\documentclass[twocolumn]{aastex62}

\usepackage{amsmath,amssymb}
\usepackage{color}
\usepackage{natbib}
\usepackage{graphicx}
\usepackage{lipsum}
\usepackage{epsfig}
\usepackage{hyperref}

\newcommand{\mr}{\mathrm}
\newcommand{\om}{\omega_\mr m}

\newcommand{\Om}{\Omega_\mr m}
\newcommand{\Omb}{\Omega_\mr b}

\renewcommand{\d}{{\rm d}}
\newcommand{\sig}{\sigma_8}
\newcommand{\ns}{n_s}
\newcommand{\w}{w_0}
\newcommand{\wa}{w_a}

\newcommand{\pd}{P_{\delta}}
\newcommand{\eps}{\epsilon}

\def\vek{\mathbf}
\newcommand{\like}{L}
\newcommand{\prob}{P}
\newcommand{\probr}{P_r}
\newcommand{\pco}{\vek p_\mr{co}}
\newcommand{\pnu}{\vek p_\mr{nu}}
\newcommand{\D}{\vek D}
\newcommand{\M}{\vek M}
\newcommand{\matC}{\mathbf C}

\newcommand{\bea}{\begin{eqnarray}}
\newcommand{\eea}{\end{eqnarray}}
\newcommand{\nn}{\nonumber}
\newcommand{\be}{\begin{equation}}
\newcommand{\ee}{\end{equation}}

\shorttitle{Kinematic Lensing} \shortauthors{Huff et al.}

\begin{document}

\title{Kinematic Lensing - Cosmic Shear Without Shape Noise}

\correspondingauthor{Eric Huff}
\email{Eric.M.Huff@jpl.nasa.gov}
\author{Eric M. Huff}
\affiliation{Jet Propulsion Laboratory, California Institute of Technology, Pasadena, CA 91109, USA}

\author{Elisabeth Krause}
\affiliation{Steward Observatory/Department of Astronomy, University of Arizona, 933 North Cherry Avenue, Tucson, AZ 85721, USA}

\author{Tim Eifler}
\affiliation{Steward Observatory/Department of Astronomy, University of Arizona, 933 North Cherry Avenue, Tucson, AZ 85721, USA}

\author{Xiao Fang}
\affiliation{Steward Observatory/Department of Astronomy, University of Arizona, 933 North Cherry Avenue, Tucson, AZ 85721, USA}

\author{Matthew R. George}
\affiliation{Department of Astronomy, University of California, Berkeley,CA 94720, USA}
\affiliation{Lawrence Berkeley National Laboratory, 1 Cyclotron Rd., Berkeley, CA 94720}

\author{David Schlegel}
\affiliation{Lawrence Berkeley National Laboratory, 1 Cyclotron Rd., Berkeley, CA 94720}
\keywords{cosmology: observations --- gravitational lensing: weak ---
  methods: observational}

\begin{abstract}
  We describe a new method for reducing the shape noise in weak lensing measurements by an order of magnitude. Our method relies on spectroscopic measurements of disk galaxy rotation and makes use of  the Tully-Fisher relation in order to control for the intrinsic  orientations of galaxy disks. For this new proposed method, so-called Kinematic Lensing (KL), the shape noise ceases to be an important source of statistical error.   
  
  We use the \textsc{CosmoLike} software package to simulate likelihood analyses for two Kinematic Lensing survey concepts (roughly similar in scale to Dark Energy Survey Task Force Stage III and Stage IV missions) and compare their constraining power to a cosmic shear survey from the Large Synoptic Survey Telescope (LSST). Our forecasts in seven-dimensional cosmological parameter space include statistical uncertainties resulting from shape noise, cosmic variance, halo sample variance, and higher-order moments of the density field. We marginalize over systematic uncertainties arising from photometric redshift errors and shear calibration biases considering both optimistic and conservative assumptions about LSST systematic errors.   
  
  We find that even the KL-Stage III is highly competitive with the optimistic LSST scenario, while evading the most important sources of theoretical and observational systematic error inherent in traditional weak lensing techniques. Furthermore, the KL technique enables a narrow-bin cosmic shear tomography approach to tightly constrain time-dependent signatures in the dark energy phenomenon.	  
\end{abstract}

\section{Introduction}{\label{sec:intro}}
Weak gravitational lensing has been advertised as a powerful probe of
cosmology (e.g. \citealt{abc06, 2008ARNPS..58...99H,wme13}), and is a
major science driver for several ongoing and future surveys, such as
the Dark Energy Survey\footnote{http://www.darkenergysurvey.org/}, the
KIlo Degree Survey\footnote{http://kids.strw.leidenuniv.nl/},
HyperSuprimeCam\footnote{http://www.naoj.org/Projects/HSC/}, the Large
Synoptic Survey Telescope \footnote{http://www.lsst.org/lsst/}, Euclid
\footnote{sci.esa.int/euclid/}, and the Wide-Field Infrared Survey
Telescope\footnote{http://wfirst.gsfc.nasa.gov/}. It is the least
indirect method available for constraining the distributions of both
dark and luminous matter in the universe. Weak lensing by large-scale
structure -- termed cosmic shear -- promises powerful constraints on
both the growth of structure and the expansion history of the
Universe.

For cosmic shear, typical fluctuations in the matter density field
projected over cosmological distances produce lensing distortions to
galaxy ellipticities of order $10^{-3}$. The noise (per ellipticity
component) resulting from the random intrinsic orientations and
ellipticities of shapes, by contrast, is $\sigma_\eps \sim0.26$
(e.g. \citealt{cjj13}). In order to detect the cosmic shear signal at high significance, lensing analyses must include faint and poorly-resolved galaxies. This comes at a high cost in increased systematic error, as shear
measurements using marginal galaxy images are
especially susceptible to calibration biases
(c.f. \citealt{2003MNRAS.343..459H,2013MNRAS.429..661M}). For all of
these reasons, it is highly desirable to control for sources of
intrinsic scatter in lensing observables. 

Several methods have been proposed for reducing the shape noise using
additional observables to infer the unlensed properties of
galaxies. Polarization in radio observations provides an estimate of
the unlensed position angle
(e.g.,\citealt{2011ApJ...735L..23B}). Spatially-resolved kinematic
maps carry information about the intrinsic orientation
\citep{2002ApJ...570L..51B,2006ApJ...650L..21M}. In the context of
weak lensing magnification, a scaling relation can be used to predict
the unlensed size of a galaxy from other photometric quantities
\citep{2006ApJ...648L..17B, heg2014}. This paper presents
a novel combination of the latter two approaches, employing
minimally-resolved disk galaxy kinematics and the Tully-Fisher scaling
relation to estimate both components of shear while suppressing shape
noise.

This idea proposed in this paper benefits from the fact that 
the coming decade is likely to see a considerable increase in the capacity of massively multi-object spectroscopy of galaxies at moderate redshifts. Two such instruments
currently under development at the time of this writing include the
Prime Focus Spectrograph for the Subaru telescope
\citep{2014PASJ...66R...1T} and the DESI spectrograph
\citep{2013arXiv1308.0847L}. The primary science surveys anticipated
for these instruments require spectroscopic target densities of $1\;
{\rm arcmin}^{-2}$, which is nearly an order of magnitude above the
previous generation of spectroscopic surveys \citep{boss, WiggleZ}.

It is the coincidence between this surge in spectroscopic capacity and
the widespread scientific interest in weak lensing that motivates the
present work. We forecast the cosmological constraining power of a
cosmic shear measurement using a large spectroscopic data set (in
combination with high-quality imaging) comparable in size to those
expected from the aforementioned multi-object spectrographs. We show
that, by using a combination of minimally-resolved disk galaxy
kinematics and the Tully-Fisher scaling relation, a spectroscopic weak
lensing experiment has the potential to greatly improve on the
statistical and systematic errors of conventional lensing
measurements.
\begin{figure}[t]
\includegraphics[width=\linewidth, bb= 150 320 540 500,clip]{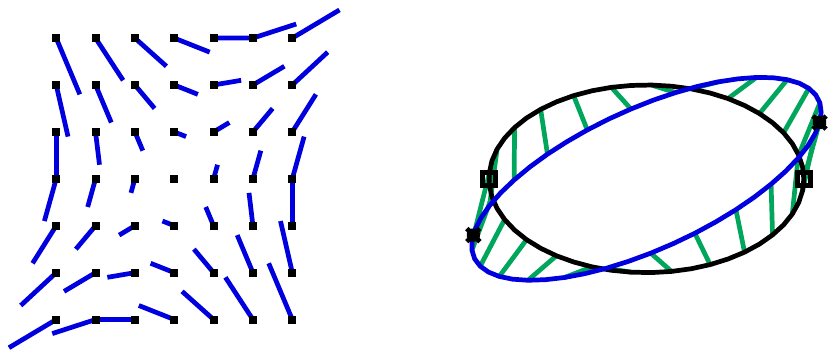}
\caption{The distortion field (left) produced by a $\gamma_{\times}=0.2$ shear, and the effect of this distortion on an ellipse (right). Note that the sheared ellipse is not a rotated version of the unsheared because the extremal points of the sheared and unsheared ellipse do not map to one another.}
\label{fig:illustration}
\end{figure}

\section{Basic Concepts}

In this section we introduce the fundamental observables and key equations necessary to relate kinematic observables to the lensing signal. We aim to show that the two components of the weak gravitational lensing shear distortion can be inferred directly from images and spectra of an individual galaxy. We note that this does require knowledge of the slope and intercept of the TFR, but this information can be retrieved by fitting the ensemble of observed galaxies.

The reason this is possible can be seen by close inspection of Figure~\ref{fig:illustration}. After the shear, the loci originally corresponding to the maximum and minimum of the line-of-sight velocity field (open squares) are no longer located along the apparent major and minor axes of the ellipse, instead appearing at the points indicated by the solid squares. The rotation speed measured along the major axis is always reduced relative to the unsheared case, and the rotation speed measured along the minor axis is always non-zero. {\it It is thus the difference between the velocity expected from the imaging and the measured velocity field that allows the shear to be inferred for individual galaxies}.

There are two components to the shear, and both can be inferred directly by combining spectroscopy and imaging. We illustrate this further with some simple analytic arguments for measuring the component of the shear aligned with the major axis of galaxy, $\gamma_+$. The shear component aligned at $\frac{\pi}{4}$ relative to the major axis, $\gamma_\times$, is treated in the more formal derivation in section~\ref{sec:formalism}. 

Consider the idealized case of an  rotationally-supported disk galaxy inclined at some angle $i$ with respect to the observer's line of sight. This galaxy's luminosity $L$ and rotation speed (as determined from spectroscopy) $v_{\rm spec}$ are related by the Tully-Fisher Relation (TFR):
\begin{align}
v_{\rm spec} = v_{\rm TF} \sin i
\label{eqn:vTF}
\end{align}
where the TFR, relating a stellar mass or absolute magnitude $M_B$ and the disk galaxy's circular velocity $v_{\rm TF}$, is commonly parameterized with slope $a$, pivot $M_p$, and intercept $b$ as
\begin{align} 
\log v_{\rm TF} = a\log (M_B + M_p) + b~.
\end{align}

The rotation speed measured along the {\it minor} axis is zero, within measurement errors.

Next, we introduce a gravitational lensing signal, applying a shear at some arbitrary angle with respect to the major axis of the galaxy image. Figure~\ref{fig:illustration} shows the result. The lensing distortion re-maps points in the image plane as shown in the left panel, inducing the change in orientation and axis ratio shown in the right panel. 

The lensing observables are typically described in terms of the ellipticity, the magnitude of which we define here as  
\begin{align}
e = \frac{1-q^2}{1+q^2}
\label{eqn:ellipdef}
\end{align}
where $0 \le q < 1$ is the semi-minor to semi-major axis ratio of the isophotes of the galaxy image.

The component of the shear aligned with the major axis of the galaxy, $\gamma_+$, transforms the intrinsic (unlensed) ellipticity $e_{\rm int}$ into the observed ellipticity $e_{\rm obs}$ as
\begin{align}
e_{\rm obs} =  e_{\rm int} + 2(1-e_{\rm int}^2) \gamma_+
\label{eqn:shear_ellip}
\end{align}
\citep{bej02}. 

With only the ellipticity and position angle, it is impossible to tell the difference between the effects of a shear and a change in $e_{\rm int}$. Wide-field cosmological lensing surveys to date thus need to rely on the fact that the shear is coherent over large length scales, while the intrinsic alignments are, to a good approximation, a shorter-range effect.

The situation changes if $v_{\rm spec}$ is measured and $v_{\rm TF}$ is known. Equation~\ref{eqn:vTF} then allows us to directly infer an inclination $\sin i$, and hence an ellipticity $e_{\rm spec}$ from the spectroscopic observables. As is typical when deriving inclination corrections in TFR studies, we assume that $e_{\rm spec} = e_{\rm int}$, which allows us to solve Equation~\ref{eqn:shear_ellip} directly for the shear. 

Note that noise in this relation contributes directly to the intrinsic scatter in the TFR. The latter is constrained by a wide range of observations \citep{2011ApJ...741..115M,2012MNRAS.425.2610R, 2016MNRAS.460..103T,2018MNRAS.tmp.2661T}, which place an upper bound on the scatter between $e_{\rm spec}$ and $e_{\rm int}$ and motivate the shape noise assumed in the forecasts we describe in section~\ref{sec:scatter}.

What follows is a detailed derivation of the formalism connecting the spectroscopic and imaging observables described above to the cosmological lensing signal.

 \subsection{The Effect of Shear on Kinematic Observables}
 \label{sec:formalism}
 In this section, we use toy models of galactic disks to demonstrate how kinemetry can break the degeneracy between shear and shape, controlling for a large fraction of the intrinsic shape noise of disk galaxies. We show that, to linear order in the shear, $\gamma_+$ changes the amplitude of the ellipticity of the galaxy image and $\gamma_\times$ produces an apparent rotation of said image. That is, $\gamma_\times$ changes the galaxy ellipse's position angle by $\theta$.

 We define the shear as a linear distortion of the image $\boldsymbol{x}'= \mathcal{A}\:\boldsymbol{x}$, where $\boldsymbol{x}=(x,y)^{\rm T}$ and $\boldsymbol{x}'=(x',y')^{\rm T}$ are the coordinates on the source plane and the image plane, respectively,
 \begin{align}
  \mathcal{A} = \left(
 \begin{array}{cc}
 1+\gamma_+ & \gamma_\times \\
 \gamma_\times & 1-\gamma_+
 \end{array}
 \right).
 \end{align}
 We will discuss the effects of lensing on a idealized, circular rotating disk galaxy, with finite edge-on aspect ratio $q_z$. We choose our coordinate system for the shear distortion such that positive $\gamma_{+}$ induces a shear along the major axis of the galaxy, which means the shape of the unlensed galaxy can be described by an ellipse
\begin{equation}
q^2 x^2+y^2 = 1~.
\end{equation}
Under the lensing transformation $\mathcal{A}$ and keeping terms up to the first order in shear components, the lensed galaxy becomes
\begin{equation}
q^2(1-4\gamma_+)x'^2+y'^2 -2(1+q^2)\gamma_\times x'y'= 1~,
\label{eq:shear_ellipse}
\end{equation}
which is a squeezed (or stretched) and rotated ellipse, comparing to the unlensed galaxy. To see this, let's define an ellipse $q_{\rm obs}^2x^2+y^2=1$, where $q_{\rm obs}$ is related to $e_{\rm obs}$ by
\begin{equation}
e_{\rm obs} = \frac{1-q_{\rm obs}^2}{1+q_{\rm obs}^2}~.
\end{equation}
We choose our coordinates such that the unlensed galaxy has position angle $\theta_{\rm int}=0$; after rotating the galaxy by the observed position angle $\theta_{\rm obs}$
\begin{equation}
\mathcal{R} = \left(\begin{array}
{cc}1&\theta_{\rm obs} \\
-\theta_{\rm obs} &1
\end{array}\right) \, ,
\end{equation}
we obtain the equation of the rotated ellipse
\begin{equation}
q_{\rm obs}^2x'^2+y'^2-2\theta_{\rm obs}(1-q_{\rm obs}^2)x'y'=1~,
\label{eq:rotat_ellipse}
\end{equation}
which matches Eq.~(\ref{eqn:shear_ellip}) with
\begin{align}
q_{\rm obs}^2 &= q^2(1-4\gamma_+)~,\\
\theta_{\rm obs}(1-q_{\rm obs}^2) &=(1+q^2)\gamma_\times~.
\end{align}
Keeping terms up to the first order in shear components and the rotation angle, we obtain
\begin{align}
e_{\rm obs} &= e_{\rm int} + 2(1-e_{\rm int}^2)\gamma_+~,\label{eqn:e_obs}\\
\theta_{\rm obs} &= \frac{\gamma_\times}{e_{\rm int}}~. 
\end{align} 
We can see that at linear order in the shear, the even- and odd-parity components have different effects on the image, but in both cases the shear is degenerate with the parameters of the unlensed ellipse. The observables ($e_{\rm obs}, \theta_{\rm obs}$) depend on both ($\gamma_+$,$\gamma_{\times}$) and ($e_{\rm int}$, $\theta_{\rm int}$).

We now show how to break this degeneracy and infer both shear components by combining the observed galaxy shapes with kinematic measurements.

For a disk galaxy with an edge-on aspect ratio $0<q_z<1$, the inclination is related to the observed axis ratio $q$ by
\begin{align}
   \sin^2 i &= \frac{1-q^2}{1-q_z^2}~.
\label{eqn:sini}
\end{align}
We can re-arrange this relation and combine it with Eq.~(\ref{eqn:vTF}) to get
\begin{equation}
e_{\rm int} = \frac{\left(1-q_z^2\right)\left(v_{\rm spec}/v_{\rm TF} \right)^2}{2 -
               \left(1-q_z^2\right) \left(v_{\rm spec}/v_{\rm TF} \right)^2}~.
\end{equation}
Knowledge of the Tully-Fisher relation and measurement of $v_{\rm spec}$ allows for determination of the intrinsic, unlensed ellipticity $e_{\rm int}$. We can now rewrite Eq.~(\ref{eqn:e_obs}) solving for $\gamma_+$ 
\begin{equation}
\gamma_+ = \frac{e_{\rm obs} - e_{\rm int} }{2\left(1-e^2_{\rm int}\right)} 
\label{eqn:gplus}
\end{equation}
and we note that the rhs is comprised of known quantities.

To sum up, a disk galaxy's line-of-sight velocity offset from the TFR predicts an ellipticity $e_{\rm int}$. The difference between this ellipticity and that of the observed image is proportional to the weak lensing shear component aligned with the galaxy's major axis $\gamma_+$.

Second, the velocity measured along the minor axis of the sheared ellipse, $v'_{\rm minor}$, informs us of the rotation angle $\theta_{\rm obs}$, as well as $\gamma_\times$. Assume that the point on the minor axis has coordinate $\boldsymbol{x'}_{\rm minor}=(\cos\vartheta'_{\rm minor},\sin\vartheta'_{\rm minor})^{\rm T}$, where $\vartheta'_{\rm minor}=\pi/2+\theta_{\rm obs}$ is the polar angle from the positive $x$-axis. Before being lensed, the point, at linear order, is located at $(\cos\vartheta,\sin\vartheta)$, i.e.,
\begin{equation}
\mathcal{A}^{-1}\boldsymbol{x'}_{\rm minor} = \left(
\begin{array}{c}
-\theta_{\rm obs} - \gamma_{\times} \\
1
\end{array}
\right)\simeq\left(\begin{array}{c}
\cos\vartheta\\
\sin\vartheta
\end{array}
\right)~,
\end{equation}
which gives $\cos\vartheta = -\theta_{\rm obs}-\gamma_\times=-\gamma_\times(1+1/e_{\rm int})$. Note that $\cos\vartheta$ is also related to $v'_{\rm minor}$ by
\begin{equation}
\frac{v'_{\rm minor}}{v_{\rm TF}} = \cos \vartheta \sin i ~,
\end{equation}
we can solve for $\gamma_\times$ as
\begin{align}
\gamma_\times &= -\frac{\cos\vartheta}{1+1/e_{\rm int}}\nonumber\\
&=-\frac{1}{1+1/e_{\rm int}}\frac{v_{\rm minor}'}{v_{\rm TF}}\sqrt{\frac{(1-q_z^2)(1+e_{\rm int})}{2e_{\rm int}}}\nonumber\\
&=-\frac{v_{\rm minor}'}{v_{\rm TF}}\sqrt{\frac{(1-q_z^2)e_{\rm int}}{2(1+e_{\rm int})}}~. \label{eqn:gcross}
\end{align}

It can be verified that the effects of $\gamma_+$ on the minor axis,
and the effects of $\gamma_\times$ on the major axis are both of
quadratic order (see \citealt{bej02}, section 2.2).

\section{A Tully-Fisher Weak Lensing Survey}
\label{sec:tfsurvey}
\subsection{Effective Shape Noise}
\label{sec:scatter}
We estimate the effective shape noise that would arise from a
hypothetical TF lensing experiment by generating catalogs of mock
observables with appropriate noise properties. Each quantity in
Eqs. \ref{eqn:gplus} and \ref{eqn:gcross} is generated according to the following procedure, with all parameters drawn from \citet{2012MNRAS.425.2610R}.:
\begin{enumerate}
\item An absolute magnitude $M_B$ for each mock catalog entry is drawn
  from a normal distribution with mean -20.5 and standard deviation of
  unity.
\item For each mock catalog entry, $\log_{10} v_{\rm circ}$ is drawn
  from a Gaussian with mean $2.142 - 0.128(M_B +20.558)$, and a
  standard deviation (modeling the intrinsic TFR scatter) of
  $\sigma_{\rm int}=0.033$. 

\item The cosine of the inclination angle $i$ is drawn uniformly from
  $[0,1)$, and the image axis ratio $q$ is assigned as per
  Eq. \ref{eqn:sini}.
\end{enumerate}
Both shear responses in Eqs.~(\ref{eqn:gplus},\ref{eqn:gcross}) are quite sensitive to the line-of-sight orientation of the galaxy in question, suggesting that there are substantial gains to be had from weighting a shear estimate accordingly. Here we calculate the effective shape noise, weighting by each galaxy's observables' shear sensitivity:
\begin{align}
R_+ &= 2(1 - e_{\rm int}^2)\\
R_{\times} &= \sqrt{\frac{2}{1-q_z^2}\frac{1+e_{\rm int}}{e_{\rm int}}}
\end{align}

We calculate the weighted standard deviation in recovered shears for both shear components from our monte-carlo draws using the above procedure, and find values of $\sigma_+ = 0.038$ and $\sigma_\times = 0.014$. We adopt the geometric average for our estimate of the effective shape noise, which yields $\sigma_{\epsilon,{\rm TF}} = 0.023$. It should be noted that this number assumes that the kinematic measurements are dominated by the intrinsic Tully-Fisher and internal disk kinematic dispersions; we defer the modeling the impact of realistic measurement effects to future work. \citealt{2015PASA...32...40D}, who attempt a similar forecast for shear dispersion using more detailed simulations of the measurement, also find few-percent level shear estimation errors.

LSST-equivalent levels of shape noise should be achievable using kinematics 
with spectra for $0.4$ galaxies per square arcminute. This is comparable to
the target densities planned for the next generation of large
spectroscopic surveys, and while the instruments currently under
construction for these surveys have not been designed to obtain the
spatially-resolved spectroscopy necessary for a spectroscopic lensing
survey, they may be capable of the measurements discussed here. We
discuss this point in general terms in Sect.~\ref{sec:instrument}, but
defer instrument-specific survey considerations to a later analysis.

\subsection{Designing a Tully-Fisher Lensing Survey}
Here we describe two TF survey concepts. The first (hereafter TF-Stage
III) is intended to be representative of an experiment that could be
performed with instruments similar to those currently under
development, relies on optical spectroscopy to measure rotation
curves, and covers $5,000$ square degrees. The second (hereafter
TF-Stage IV) is intended to represent a more optimistic future survey,
and assumes a greater redshift reach (which will require an infrared
spectrograph) and a survey area of $15,000$ square degrees, which is
similar to the planned LSST footprint after masking \citep{cjj13}.

We estimate the number and redshift distribution of viable targets for
each of these two surveys with the Cosmos Mock Catalog (CMC)
\citep{2009A&A...504..359J}. The CMC, created using data from
COSMOS\footnote{http://cosmos.astro.caltech.edu/}, zCOSMOS
\citep{2007ApJS..172...70L}, and
GOODS-N\footnote{http://www.stsci.edu/science/goods/}, was designed
specifically for tuning target selection criteria for future
wide-field imaging and spectroscopic surveys. Spectroscopic templates
were fit to the $> 500,000$ galaxies detected in COSMOS, and the
spectral template assignment and luminosity function were validated
using zCOSMOS and the deeper GOODS-N imaging, respectively. The CMC
has been updated since its original publication; we use the version
available on the project
website\footnote{http://lamwws.oamp.fr/cosmowiki/RealisticSpectroPhotCat}
as of December 2011.

\begin{figure}
\includegraphics[width=\linewidth, bb= 130 160 520 630,clip]{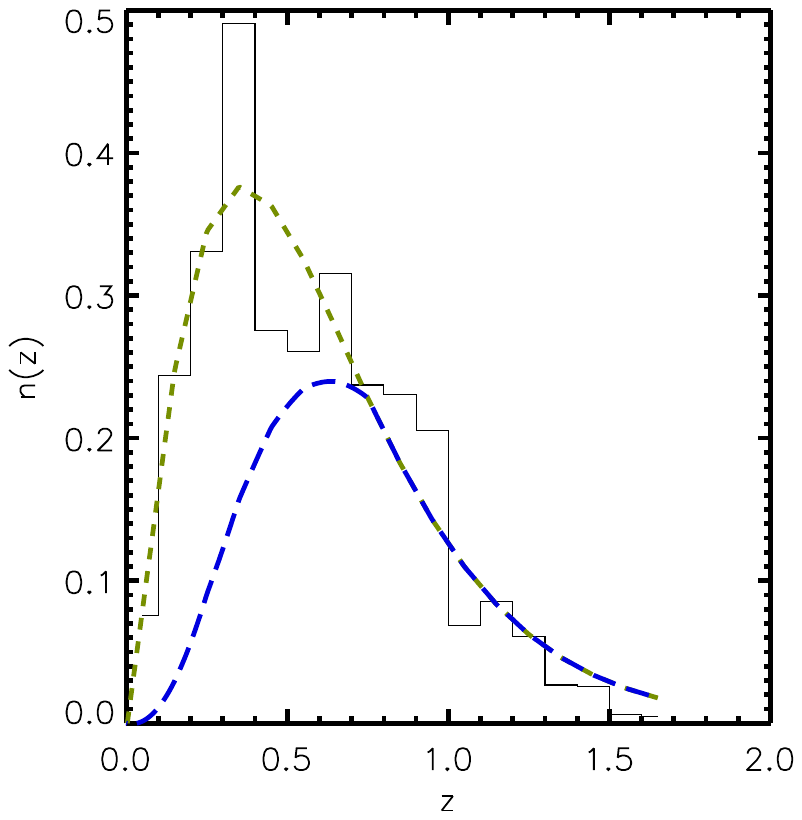}
\caption{Redshift distributions from the CMC (solid, histogram) for
  the TF-Stage III experiment. The smooth fit to this is the short-dashed
  green line, and the redshift distribution used to construct the
  TF-Stage III covariances is shown in the long-dashed blue line. }
\label{fig:nz_cmc}
\end{figure}

For TF-Stage III, we define a viable TF target as one that meets the following criteria:
\begin{enumerate}
  \item half light radius $\geq 0.5''$
  \item $r$-band magnitude $\leq$ 23.5
  \item morphological template type $>8$ (this excludes ellipticals and S0 galaxies)
  \item $1,000\,{\rm\AA} \leq \lambda_{\rm line} \leq 10,000\,{\rm \AA}$
  \item line emission line flux $\geq 10^{-16}\:{\rm erg \:s^{-1}\:cm^{-2}}$
\end{enumerate}
We require that at at least one of the $\left[\rm{OII}\right]$,
$\left[\rm{OIII}\right]$, ${\rm H}\alpha$, or ${\rm H}\beta$ emission
lines meet both of these criteria. For TF-Stage IV, we extend the
spectroscopic window to $20,000{\rm\AA}$.

The first two requirements permit detection and shape measurement from
photometric catalogs. The third limits the sample to disks, and the
fourth to objects with line emission -- specifically, line emission
that traces the gas disk -- in a wavelength range accessible to
ground-based spectroscopy.  The final requirement ensures that the
line emission at 2.2 disk scale lengths be above the typical sky
background at $8,000\,{\rm \AA}$ of $10^{-17}{\rm erg\: s^{-1}}$, and
is motivated by previous studies \citep{2011ApJ...741..115M} which
find that rotation curve measurements are most reliable when the
emission line is detected out to this distance from the galaxy
center. The actual line emission detection threshold will of course
depend on the exposure time. Achieving this signal-to-noise ratio
should be possible on an $8-$m telescope with a PFS-like spectrograph
in $30$-minute exposures\footnote{as we are targeting larger, brighter
  galaxies than the PFS and DESI surveys, the fractional contribution
  of sky flux to the total flux in each fiber is substantially smaller
  than for the redshift survey components of those programs.}, which
for this program would entail approximately 4-5 years of dedicated
observations.

The available galaxy density set by applying these constraints to the
CMC is $2.9/{\rm arcmin}^{2}$. This number is most sensitive to the
emission line strength requirement; halving the emission line
detection threshold approximately doubles the available target
density. 

We do not expect a feasible spectroscopic lensing survey to
realistically exceed a target density of one galaxy per square
arcminute. To construct the redshift distributions for both TF
surveys, we first fit a smooth distribution of the usual form:
\begin{equation}
p(z) \propto z^{\alpha} e^{-(\frac{z}{z_0})^{\beta}}
\end{equation}
to the redshift distribution of CMC sources that meet the selection
criteria described above. We then subsample this to our fiducial
target density assuming that the high-redshift tail is left in place,
smoothly reducing the number density at lower redshift in a manner
proportional to the comoving volume. The resulting redshift
distributions for the CMC selection, its smoothed fit, and the
fiducial survey redshift distributions for the TF-Stage III experiment are
shown in Fig.~\ref{fig:nz_cmc}.

\subsubsection{Instrumental Prospects}
\label{sec:instrument}
This paper argues that a TF lensing survey can produce cosmological
constraints competitive with other Stage IV dark energy experiments
with a sufficient number of spatially resolved disk galaxy
spectra. Several wide-field imaging surveys (HSC, LSST, DES) with a
weak lensing focus are already planned or underway; we assume that any
of these might be used for target selection and shape measurement
for a TF survey. The primary obstacle is the collection of order
$10^7$ resolved spectra. 

Two massively multi-object fiber-fed spectroscopic instruments are
currently in the advanced planning stage: the Prime Focus Spectrograph
for the Subaru telescope \citep{2014PASJ...66R...1T} and the DESI
spectrograph \citep{2009arXiv0904.0468S}. Each is capable of producing
target densities in a single exposure of $0.5$ per square
arcminute. Spatially resolved spectroscopy can in principle be
obtained with multiple pointings. DESI, in particular, is planning to
collect $50$ million galaxy spectra, the majority of which are at
$z>1$. While we defer a more detailed, instrument-specific feasibility
study to a future paper, it seems clear that a TF lensing survey is
not drastically more challenging than currently planned projects.

\section{Modeling Cosmological Quantities}
\label{sec:model}
We present a side-by-side comparison of a Stage IV Dark Energy
experiment (pseudo-LSST) and the proposed Tully-Fisher measurements. In
this section, we present a calculation of the expected cosmological
constraints from each of these two surveys, including both
statistical and systematic error contributions. The following sections
describe the prediction code, the systematic errors we consider here,
and our method for incorporating the systematics into our model.

\begin{figure}[t]
\includegraphics[width=\linewidth, bb= 130 160 500 630,clip]{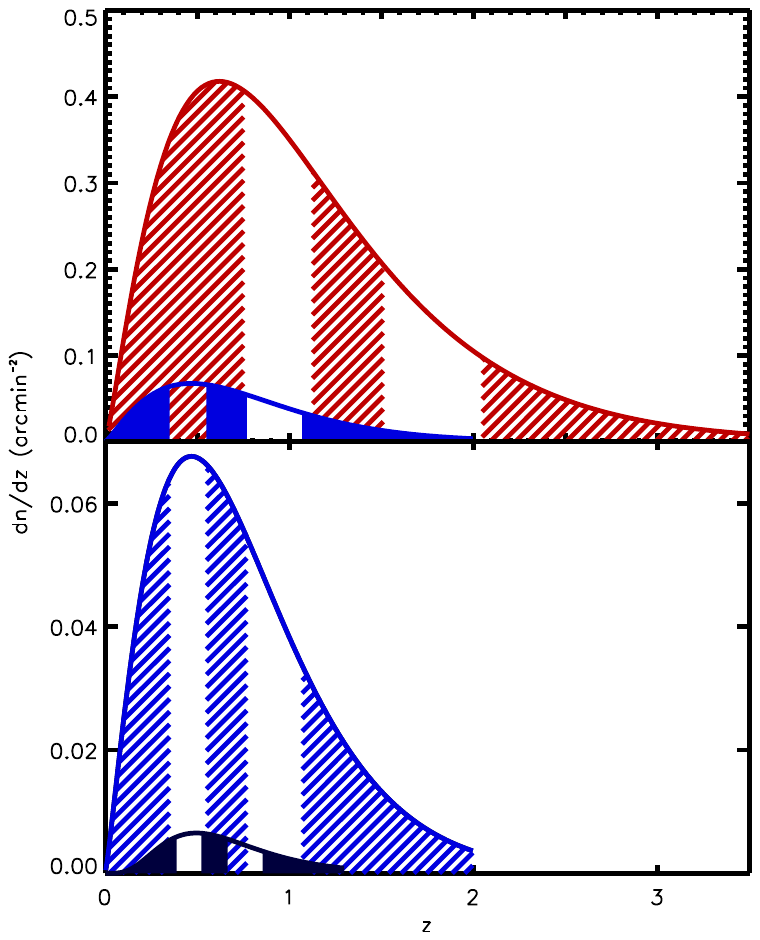}
\caption{Redshift distributions for our model LSST, DES, and
  fiducial Tully-Fisher redshift surveys. The top panel shows projections for LSST (red,
  with line fill) and DES (blue, solid fill). The bottom panel shows
  DES (blue, line fill) and the TF-Stage III survey (black, solid fill).}
\label{fig:nz_comparison}
\end{figure}

\subsection{Prediction Code}
\label{sec:pred}
The simulated likelihood analysis in this paper is computed using the cosmic shear module of
\textsc{CosmoLike} \citep{kre17}, which has been used in several ongoing data analysis as well as forecasting efforts of future surveys \citep{eks14,ekd15,keb16,ske17,kez17,DES17,2018arXiv180403628D}. We use the fastest version of \textsc{CosmoLike}, which computes the linear power spectrum using the \cite{eh99} transfer function and model
the non-linear evolution of the density field as described in
\cite{tsn12}. We compute time-dependent dark energy models
($w=w_0+(1-a)w_a$) following the recipe of {\sc icosmo} \citep{rak11}, which in the non-linear regime interpolates Halofit between flat and open cosmological models \citep[please also see][for more details]{shj10}.\\
From the density power spectrum we compute the shear power spectrum as
\begin{equation}
\label{eq:pdeltatopkappa}
C ^{ij} (l) = \frac{9H_0^4 \Om^2}{4c^4} \int_0^{\chi_\mr h} 
\mr d \chi \, \frac{g^{i}(\chi) g^{j}(\chi)}{a^2(\chi)} \pd \left(\frac{l}{f_K(\chi)},\chi \right) \,,
\end{equation}
with $l$ being the 2D wave vector perpendicular to the line of sight,
$\chi$ denoting the comoving coordinate, $\chi_\mr h$ is the comoving
coordinate of the horizon, $a(\chi)$ is the scale factor, and
$f_K(\chi)$ the comoving angular diameter distance.\\
The lens efficiency $g^{i}$ is defined as an integral over the
redshift distribution of source galaxies $n(\chi(z))$ in the
$i^\mr{th}$ tomographic interval
\begin{equation}
\label{eq:redshift_distri}
g^{i}(\chi) = \int_\chi^{\chi_{\mr h}} \mr d \chi' n^{i} (\chi') \frac{f_K (\chi'-\chi)}{f_K (\chi')} \,.
\end{equation}
Since we chose five tomographic bins, the resulting data vector which
enters the likelihood analysis consists of 15 tomographic shear power
spectra, each with 20 logarithmically spaced bins ($l \in [30;5000]$), hence 300 data points
overall. In the following analysis we assume different redshift
distributions (depending on the probe/survey considered), however we
always choose five tomography bins with equal number densities in each
z-bin.

\subsection{Statistical Covariances}
\label{sec:cov}
Under the assumption that the shear field is Gaussian (which means that the shear 4pt-function can be expressed in terms of 2pt-functions) the covariance of projected shear power spectra can be expressed as \citep{huj04} 

\begin{widetext}
\be
\label{eq:covhujain}
\mr{Cov_G} \left( C^{ij} (l_1) C^{kl} (l_2) \right) = \langle \Delta C^{ij} (l_1) \, \Delta C^{kl} (l_2) \rangle  =  \frac{2 \pi \, \delta_{l_1 l_2}}{A l_1 \Delta l_1}  \left[\bar C^{ik}(l_1) \bar C^{jl}(l_1) + \bar C^{il}(l_1) \bar C^{jk} (l_1) \right]\,,
\ee
\end{widetext}

with
\be
\label{details}
\bar C^{ij}(l_1)= C^{ij}(l_1)+ \delta_{ij} \frac{\sigma_\eps^2}{n^{i}} \,,
\ee
where the superscripts indicate the redshift bin and $n^{i}$ is the
density of source galaxies in the $i^\mr{th}$ redshift bin.

Since non-linear structure growth at late time induces significant
non-Gaussianities in the shear field Eq. \ref{eq:covhujain}
underestimates the error on cosmological parameters and needs to be
amended by an additional term,
i.e. $\mr{Cov}=\mr{Cov_G}+\mr{Cov_{NG}}$.  The non-Gaussian covariance
is calculated from the convergence trispectrum $T_{\kappa}$
\citep{CH01,taj09}, and we include a sample variance term
$T_{\kappa,\rm{HSV}}$ which describes scatter in power spectrum
measurements due to large scale density modes \citep{tb07, sht09},
\begin{widetext}
\be
 \mr{Cov_{NG}}(C^{ij}(l_1),C^{kl}(l_2)) =  \int_{|\mathbf l|\in l_1}\frac{d^2\mathbf l}{A(l_1)}\int_{|\mathbf l'|\in l_2}\frac{d^2\mathbf l'}{A(l_2)} \left[\frac{1}{\Omega_{\mr s}}T_{\kappa,0}^{ijkl}(\mathbf l,-\mathbf l,\mathbf l',-\mathbf l') + T_{\kappa,\rm{HSV}}^{ijkl}(\mathbf l,-\mathbf l,\mathbf l',-\mathbf l') \right] \,.
\ee
The convergence trispectrum $T_{\kappa,0}^{ijkl}$ is, in the absence of finite volume effects, defined as  
\be
\label{eq:tri2}
T_{\kappa,0}^{ijkl} (\mathbf l_1,\mathbf l_2,\mathbf l_3,\mathbf l_4) = \left( \frac{3}{2} \frac{H_0^2}{c^2} \om \right)^{4} \int_0^{\chi_h} \d \chi \, \left( \frac{\chi}{a(\chi)}\right)^4  g^i g^j g^k g^l \times \chi^{-6} \, T_{\delta,0}  \left( \frac{\mathbf l_1}{\chi}, \frac{\mathbf l_2}{\chi}, \frac{\mathbf l_3}{\chi}, \frac{\mathbf l_4}{\chi}, z(\chi) \right) \,,
\ee
with $T_{\delta,0}$ the matter trispectrum (again, not including finite volume effects), and where we abbreviated $g^i=g^i(\chi)$.\\
\end{widetext}
We model the matter trispectrum using the halo model \citep{Seljak00, CS02}, which assumes that all matter is bound in virialized structures that are modeled as biased tracers of the density field. Within this model the statistics of the density field can be described by the dark matter distribution within halos on small scales, and is dominated by the clustering properties of halos and their abundance on large scales. In this model, the trispectrum splits into five terms describing the 4-point correlation within one halo (the \emph{one-halo} term $T^{\mr{1h}}$), between 2 to 4 halos (\emph{two-, three-, four-halo} term), and a so-called halo sample variance term $T_{\mr{HSV}}$, caused by fluctuations in the number of massive halos within the survey area,
\be
\label{eq:t}
T = T_0 + T_{\mr{HSV}} = \left[T_{\mr{1h}}+T_{\mr{2h}}+T_{\mr{3h}}+T_{\mr{4h}}\right]+T_{\mr{HSV}}\;.
\ee
The \emph{two-halo} term is split into two parts, representing correlations between two or three points in the first halo and two or one point in the second halo. As halos are the building blocks of the density field in the halo approach, we need to choose models for their internal structure, abundance and clustering in order to build a model for the trispectrum. Our implementation of the one-, two- and four-halo term contributions to the matter trispectrum follows \citet{CH01}, and we neglect the three-halo term as it is subdominant compared to the other terms at the scales of interest for this analysis. Specifically, we assume NFW halo profiles \citep{NFW} with the \citet{Bullock01} fitting formula for the halo mass--concentration relation $c(M,z)$, and the \citet{ST99} fit functions for the halo mass function $\frac{ dn}{dM}$ and linear halo bias $b(M)$, neglecting terms involving higher order halo biasing.\\Within the halo model framework, the halo sample variance term is described by the change of the number of massive halos within the survey area due to survey-scale density modes; following \citet{sht09} it is calculated as
\begin{widetext}
\bea
T_{\kappa,\rm{HSV}}^{ijkl}(\mathbf l_1,-\mathbf l_1,\mathbf l_2,-\mathbf l_2)= \left(\frac{3}{2}\frac{H_0^2}{c^2}\Omega_{\mr m}\right)^4 &\times&  \int_0^{\chi_\mr h}d\chi \left(\frac{d^2 V}{d\chi d\Omega}\right)^2 \left(\frac{\chi}{a(\chi)}\right)^4 g^i g^j g^k g^l \nn \\
&\times&  \int d M \frac{d n}{d M} b(M)\left(\frac{M}{\bar{\rho}}\right)^2 |\tilde{u}(l_1/\chi, c(M,z(\chi))|^2 \nn \\
 &\times& \int d M' \frac{d n}{d M'} b(M')\left(\frac{M'}{\bar{\rho}}\right)^2 |\tilde{u}(l_2/\chi, c(M',z(\chi))|^2 \nn \\
 &\times&  \int_0^\infty \frac{k dk}{2\pi}P_\delta^{\mr{lin}}(k,z(\chi))|\tilde W(k\chi \Theta_{\mr s})|^2 \,.
\eea
\end{widetext}
\section{Simulated Likelihood Analyses}
\label{sec:simlike}
\textsc{CosmoLike} computes the analytic covariance and the data vector from a fiducial cosmology (see Table~\ref{tab:cosmology}) as described in Sect.~\ref{sec:model}. We assume the covariance to be known, implying that it is fixed with respect to cosmological parameters. This choice can influence cosmological constraints \citep{esh09}; however given that we sample a relatively limited parameter space, especially for our most important comparison (LSST optimistic vs. TF-Stage IV), we believe that it will not change our results qualitatively. We point out that data analyses from the high precision Stage IV surveys require an improved handling of theoretical uncertainties \citep[e.g.,][]{krh10}; however since the data vector is created internally in \textsc{CosmoLike} we can exclude these terms in the data and model vector.

\begin{table}[htp]
\caption{Fiducial cosmology and range of cosmological parameters used in the likelihood analyses}
\begin{center}
\begin{tabular}{|l|c c c c c c c|}
\hline
\hline
 &$\Om$ & $\sig$ & $\ns$ & $\w$ & $\wa$ & $\Omb$ & $h_0$ \\
\hline
Fiducial & 0.315 & 0.829 & 0.9603 & -1.0 & 0.0 & 0.049 & 0.673\\
Min & 0.1 & 0.6 & 0.85 & -2.0 & -2.5 & 0.04 & 0.6\\
Max & 0.6 & 0.95 & 1.06 & 0.0 & 2.5 & 0.055 & 0.76\\
\hline
\hline
\end{tabular}
\end{center}
\label{tab:cosmology}
\end{table}
In the simulated analysis we sample a seven dimensional cosmological parameter space with flat priors at the boundaries of the parameter range (see Table \ref{tab:cosmology}). We compare four different surveys (see Table~\ref{tab:survey} for the exact parameters); two purely photometric surveys mimicking DES and LSST and two versions of the Tully Fisher Lensing surveys, TF-Stage III and TF-Stage IV. For LSST we additionally consider an optimistic and a conservative systematics scenario.\\
The design of the Tully Fisher Lensing surveys is detailed in Sect.~\ref{sec:tfsurvey}; the number density of galaxies for TF-Stage III and TF-Stage IV ($1.1/\mr{arcmin}^2$) is limited by the number of spectra that can be acquired.\\
The survey parameters that we assume in the analyses are summarized in
Table~\ref{tab:survey}.
\begin{table}[htp]
\caption{Survey parameters}
\begin{center}
\begin{tabular}{|l|c c c c c c|}
\hline
\hline
Survey & area $[\mr{deg^2}]$ & $\sigma_\epsilon$ &  $n_\mr{gal}$ &$z_\mr{max}$ & $z_\mr{mean}$ & $z_\mr{med}$ \\
\hline
TF-Stage III & $5,000 $ & $0.021$ & $1.1$ & 1.68 &0.90 & 0.73\\
TF-Stage IV & $15,000 $ & $0.021$ & $1.1$ & 3.85& 1.09 & 0.84 \\
DES\footnote{Values taken from DES documents and internal communication within the DES collaboration.} & $5,000 $ & $0.26$  & $10$ & 2.0 & 0.84 &0.63\\
LSST\footnote{Values match specifications outlined in \cite{cjj13}.} & $15,000$ & $0.26$ &$31$ & 3.5 & 1.37 & 0.93 \\
\hline
\hline
\end{tabular}
\end{center}
\label{tab:survey}
\end{table}
Please note that throughout the paper $\sigma_\epsilon$ refers to the shape noise per component of the ellipticity.
\subsection{Systematic Uncertainties}
\label{sec:sys}
In addition to the seven cosmological parameters we consider up to
seven parameters for photo-z and shear calibration uncertainties. Note
that for LSST we consider two different scenarios, termed conservative
and optimistic, which differ in the range of photo-z and shear
calibration uncertainty prior. The LSST optimistic scenario assumes
major breakthroughs in photo-z and shape measurement methods compared
to the current state of the art, while the conservative scenario only
assumes modest progress.
\subsubsection{Photometric Redshifts}
\label{sec:photo-z}
In a photometric survey, galaxies are grouped into tomographic bins by
their photometric redshifts $z_{\rm{phot}}$. To account for the
degradation due to uncertainties in the photometric redshift
estimates, we compute the true underlying redshift distribution
$n_i(z)$ of galaxies in tomography bin $ z_{\rm{ph}}^{i}<
z_{\rm{ph}}<z_{\rm{ph}}^{i+1}$ as
\begin{equation}
n^i(z) = \int_{z_{\rm{ph}}^{i}}^{z_{\rm{ph}}^{i+1}} dz_{\rm{ph}}\, n(z) p(z_{\rm{ph}}|z)
\end{equation}
using a simple parameterization from \citet{2006ApJ...636...21M} to model $p(z_{\rm{ph}}|z)$, the distribution of photometric redshifts given true redshift,
\begin{equation}
p(z_{\rm{ph}}|z)= \frac{1}{\sqrt{2\pi}\sigma_z(z)}\exp\left[-\frac{\left(z-z_{\rm{ph}}-z_{\rm{bias}}(z)\right)^2}{2\sigma_z^2(z)}\right]\,
\end{equation}
i.e.  a Gaussian distribution with rms $\sigma_z$ and offset
$z_{\rm{bias}}$ from the true redshift (but see
\citealt{2010ApJ...720.1351H} for discussion of critical outliers).
We assume photometric redshift estimates to be unbiased on average
($\langle z_{\rm{bias}}\rangle=0$) and marginalize over the uncertainty
of the width of the distribution $\Delta \sigma_z$ and the uncertainty
of the redshift bias $\Delta z_{\rm{bias}}$ assuming Gaussian
distributions with parameter values listed in Table~\ref{tab:sys}.

Since the TF surveys require spectra from each galaxy
we assume no error from redshift uncertainty for these.
\subsubsection{Shear Calibration Biases}
\label{sec:calibration}
In addition to photo-z uncertainties we consider multiplicative shear
calibration bias in the analyses, which we implement as prefactors of
the modeled shear power spectra, i.e. $\mathcal M^i \, \mathcal M^j
\, C^{ij}(l)$. The superscripts $i,j$ correspond to the tomography
bins. We model shear calibration uncertainties as a Gaussian PDF
around a fiducial value of 1 and we further assume that the PDFs vary
independently in each tomography bin; hence we use five additional
parameters to model shear calibration (see Table~\ref{tab:sys} for
parameter ranges).
\begin{table}[htp]
\caption{Systematic error uncertainty parameters}
\begin{center}
\begin{tabular}{|l|r r r r|}
\hline
\hline
Survey & $\sigma_z$& $\Delta \sigma_z$ &  $\Delta z_{\rm{bias}}$ & $\Delta \mathcal M$  \\
\hline
TF-Stage III & - & - & - & $0.0032$  \\
TF-Stage IV & - & - & - & $0.0016$ \\
DES & $0.1 (1+z)$ & $0.1 \times \sigma_z $ &$0.01 $ & 0.02 \\
LSST, conservative & $0.05 (1+z)$ & $0.01$ &$0.01$  & 0.01 \\
LSST, optimistic\footnote{see http://lsst.org/files/docs/Phot-z-plan.pdf for photo-z uncertainties} & $0.05 (1+z)$& $0.002 $ &$0.003  $ & 0.002\\
\hline
\hline
\end{tabular}
\end{center}
\label{tab:sys}
\end{table}
Shear calibration uncertainty affects both photometric and Tully
Fisher Lensing surveys; however in the latter case our galaxy sample
generally has significantly higher S/N ($\mr{S/N} \ge 50$). Predicting
future progress in shear calibration performance is of course
difficult; for current measurements, however, the dominant systematic
calibration errors appear to arise from noise rectification bias,
which scales as $(\mr{S/N})^2$ \citep{2012MNRAS.425.1951R}. It seems safe
to assume that calibration biases will be reduced by limiting the
measurement to bright, well-resolved galaxies, and so we adopt the
aforementioned S/N scaling and assume a reduction in $\Delta
\mathcal M$ by a factor of $6.25$ when going from DES/LSST to the TF
experiments. We note that for TF-Stage IV we rescale the conservative LSST
shear calibration uncertainty, not the optimistic one (see Table~\ref{tab:sys}).
\subsection{Details of the Analyses and Results}
\label{sec:res}
\begin{figure*}[t]
\includegraphics[width=19cm, bb= 25 100 680 700,clip]{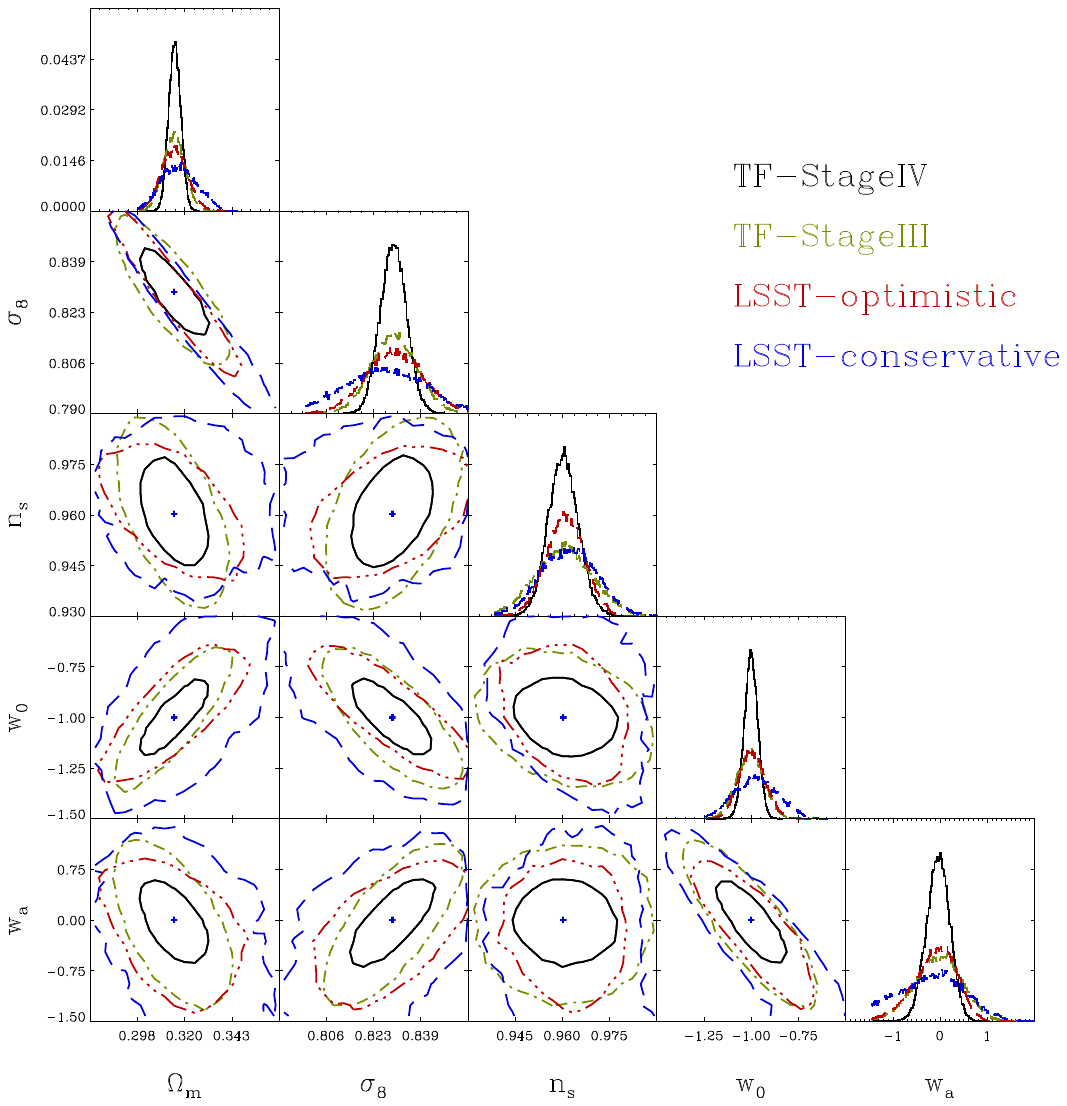}
\caption{Results of the simulated likelihood analyses. We show the
  $95\%$ confidence regions for the TF-Stage III survey (black,
  solid), the TF-Stage IV survey (red, dashed) in comparison with the
  LSST-optimistic (green, dotted) and pessimistic scenario (blue,
  dotted-dashed). We marginalize over shear calibration and (for LSST
  only) photometric redshift systematic errors.}
\label{fig:contour1}
\end{figure*}
Given the data vector and the covariance \textsc{CosmoLike} samples
the parameter space using parallel MCMC \citep{gow10} implemented
through the \textsc{emcee} python
package\footnote{http://dan.iel.fm/emcee/}. The computing time for the
300-dimensional model vector (including photo-z and multiplicative
shear calibration) at each point in parameter space is $\sim1s$, which
in combination with the parallel MCMC technique allows for an
extremely fast sampling of the considered parameter space.

We assume a Multivariate Gaussian being the functional form of the
likelihood $\like$; its width being solely determined by the
covariance matrix \be
\label{eq:like}
\like (\D| \pco, \pnu) \sim \exp \biggl( -\frac{1}{2}
\underbrace{\left[ (\D -\M)^t \, \matC^{-1} \, (\D-\M)
  \right]}_{\chi^2(\pco, \pnu)} \biggr), \ee
where $\pco$ denotes the cosmological parameter vector, $\pnu$ the
nuisance parameter vector, $\D=\D(\pco^\mr{fid},\pnu^\mr{fid})$ is the
data vector consisting of the 300 $C^{ij}(l)$ that are computed from
the fiducial model, $\M=\M(\pco,\pnu)$ is the corresponding model
vector at a given point in cosmological and nuisance parameter space,
and $\matC$ is the covariance described in Sect.~\ref{sec:cov}.

We use Bayes theorem to compute the posterior probability \be
\label{eq:bayes}
\prob(\pco, \pnu|\D) = \frac{\probr (\pco, \pnu) \,\like (\D| \pco, \pnu)}{E (\D)}
\ee
with $E$ being a normalization called evidence. We assume a flat prior probability $\probr$ in the cosmological parameter space (see Table~\ref{tab:cosmology} for details) and Gaussian priors for our nuisance parameters (see Sect.~\ref{sec:sys}). For the LSST and the Tully Fisher analyses the priors do not impact the contours at all; for the DES analysis the prior on $\wa$ cuts off outer regions of the corresponding parameter space.

Constraints that are  marginalized over nuisance parameters (or cosmological parameters that are not of interest) are calculated as
\be
\label{eq:likemarg}
L (\D|\pco ) = \int \d \vek{ \pnu} \, \exp \biggl( -\frac{1}{2}
\chi^2(\pco, \pnu) \biggr) \,.  \ee For the final runs of the
simulated likelihood analyses we compute 420,000 steps in the MCMC and
reject the first 10,000 steps as a burn-in phase. We also run several
shorter chains to check for convergence.

We produce a compressed summary of the different experiments by
computing a measure of the cosmological information content equal to
$||\Xi||^{-\frac{1}{n}}$, where $n$ is the number of cosmological
  parameters of interest, and $\Xi$ is the covariance matrix of the
  MCMC outputs
\begin{equation}
  \Xi_{ij} = {\rm cov}({p_{\rm co}}_i, {p_{\rm co}}_j) \,.
\end{equation}
This information measure corresponds roughly to the geometric average
of the constraints on the ${\pco}$, or the square of the size of the
ball in parameter hyper-space enclosing the $1-\sigma$ likelihood
surface; it is worth noting that this particular measure of
experimental merit is insensitive to the number of
parameters. Table~\ref{tab:results} shows the ratio of this quantity
for each of the four surveys considered here to that of the Dark
Energy Survey.

For better illustration we also show two-dimensional contour plots for
the most interesting cases, i.e. TF-Stage III and TF-Stage IV vs. LSST conservative
and optimistic (Fig.~\ref{fig:contour1}). These $95\%$ confidence
regions are marginalized over all other cosmological parameters (five)
and nuisance parameters (five and seven for TF and LSST, respectively).\\

\begin{figure*}[t]
\includegraphics[width=9cm]{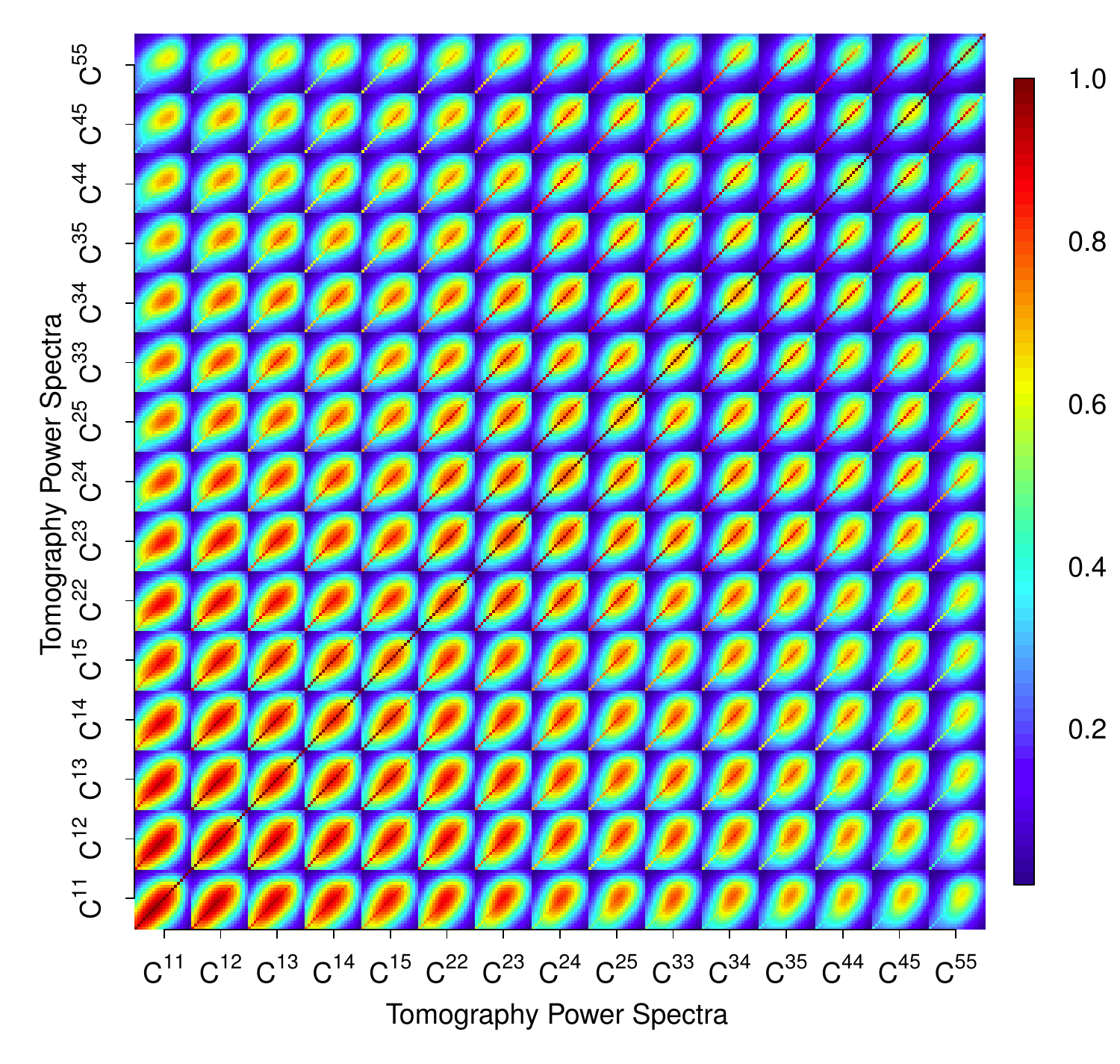}
\includegraphics[width=9cm]{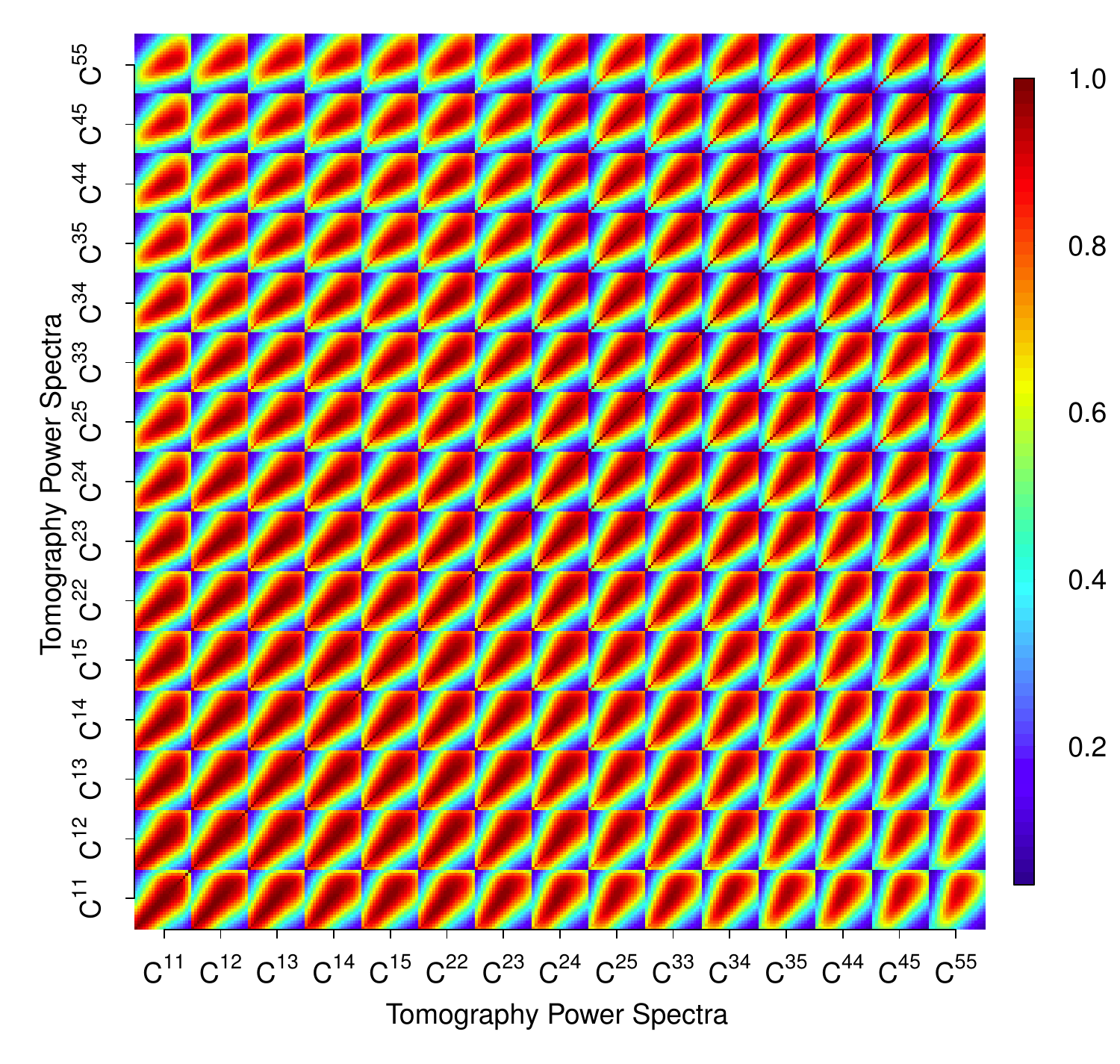}
\caption{Correlation plots of the LSST covariance (\textit{left}) and the TF covariance (\textit{right}). The data vector consists of 15 tomography power spectra, each with 20 $l$-bins, i.e. 300 data points altogether. Shape noise only acts on the main and secondary diagonals; see text for further explanation. }
\label{fig:contour2}
\end{figure*}

\begin{table*}
\caption{Marginalized One-Parameter Constraints}
\begin{center}
\begin{tabular}{|l|c c c c c c|}
\hline
\hline
Survey & $\Omega_m$ & $\sigma_8$ & $n_s$ & $w_0$, & $w_a$ & $H_0$ \\
\hline
LSST-optimistic & 0.016 & 0.015 & 0.0097 & 0.14 & 0.52 & 0.030 \\
LSST-conservative & 0.020 & 0.018 & 0.012 & 0.25 & 0.90 & 0.036 \\
TF-Stage III & 0.0011 & 0.0097 & 0.012 & 0.14 &  0.50 & 0.037 \\
TF-Stage IV & 0.0064 & 0.0056 & 0.0065 & 0.073 & 0.25 & 0.026 \\
\hline
\hline
\end{tabular}
\end{center}
\end{table*}

\begin{table}[htp]
\caption{Cosmological Information Content relative to DES}
\begin{center}
\begin{tabular}{|l|c|}
\hline
\hline
Survey & Information Content (relative to DES) \\
\hline
LSST-pessimistic & $2.39$ \\
LSST-optimistic & $3.31$ \\
TF-Stage III & $3.60$\\
TF-Stage IV & $7.10$ \\
\hline
\hline
\end{tabular}
\end{center}
\label{tab:results}
\end{table}

\subsection{Discussion}
\label{sec:discussion}

This analysis makes a number of conservative assumptions which favor
photometric weak lensing measurements. We do not include any intrinsic
alignment contamination in the LSST or DES cosmic shear forecasts, nor
do we allow for the possibility of catastrophic photometric redshift
errors. Neither of these effects are present for the TF survey
concepts, though both are important limitations of traditional methods
\citep{his04,2010ApJ...720.1351H}. We also assume that the shear
calibration biases scale as $({\rm S/N})^2$, despite claims that noise
rectification bias, which appear to be the dominant source of shear
calibration problems for many existing shape measurement methods, can
be removed at this order \citep{2012MNRAS.427.2711K}. Allowing for a
higher order calibration-${\rm S/N}$ scaling would further enhance the
cosmological information content of the TF surveys. Finally, we have
made no attempt to optimize the extraction of 3D lensing
information. It is likely that a tomographic analysis using additional
redshift bins would further improve the power of the TF-Stage III and
TF-Stage IV analyses. 

Nevertheless, Fig.~\ref{fig:contour1} and Table~\ref{tab:results} show
that the TF-Stage III experiment -- which would require only an
overlapping DESI-like spectrograph and a DES-like imaging survey -- is
comparable in constraining power to our optimistic LSST forecasts. The
TF-Stage IV experiment provides constraining power well in excess of
any other optical ground-based lensing measurement and offers a way
to break through the information ceiling set for traditional lensing
experiments by the surface density of galaxies suitable for shape
measurement.

For a better understanding of the individual error contributions we show the correlation matrices of the LSST survey (\textit{left}) and the TF-Stage III survey (\textit{right}) in Fig.~\ref{fig:contour2}. As described in Sect. \ref{sec:cov} our covariance consists of shape noise, cosmic variance (including higher order terms), and halo sample variance. Shape noise and second order cosmic variance act on the diagonal and secondary diagonal only, while halo sample variance and higher order cosmic variance act on all elements of the covariance.

For the LSST survey one can see that the larger shape noise on main and secondary diagonals dominates the submatrices at higher redshift bins and plays an important role for the low-z submatrices as well. Elements that are far from the (secondary) diagonal quickly fall off and become subdominant. In contrast, and as a result of the decreased shape noise term, the TF survey's error budget is clearly dominated by higher order cosmic variance and halo sample variance, as indicated by the large off-diagonal terms.

This dominance permits an analysis with narrower
tomographic bins, which is an extremely powerful tool to explore
time-dependent signatures in the dark energy phenomenon and separately
constrain expansion history and structure growth.

Regarding the robustness of our TF constraints we point out that an error of the disk circular velocity of 13 km/s is likely a conservative assumption. If instead we assume an error of 10 km/s, which is reasonably achievable with today's instruments already, we can tolerate a $\sim35\%$ percent failure rate in obtaining the required galaxy spectra while still achieving the constraints shown in Fig. ~\ref{fig:contour1}.

\section{Conclusions}
\label{sec:conc}
In this paper we have presented a new method to extract cosmological
information from weak gravitational lensing. Using the
well-established Tully-Fisher scaling relation we substantially
decrease the ellipticity dispersion contribution to the error budget
(from $ \sim0.26$ to $\sim0.02$), which is a major limitation of
current cosmic shear surveys. To overcome this limitation cosmic shear
surveys have to increase the depth of the survey and thereby the
number density of galaxies, which in return causes increased photo-z
and shear calibration uncertainty associated with the low S/N of
these faint galaxies.

The limitation of our method clearly is the need for spectroscopic
information and hence the limited number of galaxies that can be
observed spectroscopically within a given time interval. As a result
the model surveys we present in this paper (TF-Stage III and TF-Stage IV) have an
average number density of galaxies of $1.1/\mr{arcmin^2}$, however
they are not affected by photo-z uncertainty and only little by shear
calibration errors. For our model surveys we adopt the DETF
``Stage X'' terminology in the sense that TF-Stage III  covers $5,000
\mr{deg^2}$ (similar to DES), and TF-Stage IV  covers $15,000 \mr{deg^2}$
(similar to LSST); also we assume a similar improvement in survey
depth when going from Stage III to Stage IV.

Using the cosmic shear module of \textsc{CosmoLike} we run various
simulated cosmic shear tomography likelihood analyses, in a
multi-dimensional parameter space (seven cosmological parameters, and
five and two parameters for shear calibration and photo-z errors,
respectively). These simulated analyses include full Non-Gaussian
covariances and a realistic sampling of the parameter space which is a
major improvement over Fisher forecasts.

Our main findings are that already TF-Stage III is competitive with LSST,
depending on the assumptions of how strongly nuisance parameters
impact the LSST constraints; TF-Stage IV clearly outperforms even the
optimistic scenarios for LSST. In this context we mention all LSST
analyses assume zero intrinsic alignment contamination, which
potentially is of similar importance as photo-z and shear calibration
uncertainties.

It is however important to note that any TF-survey obviously relies on
overlapping photometric and spectroscopic data, hence the main
intention of this comparison is to strongly advocate a spectroscopic
survey overlapping with LSST. The interesting prospect of this overlap
is not just a TF-Stage IV survey but a combination of TF-Stage IV and
LSST. Galaxies without spectra will substantially contribute to the
constraints, especially since the overlap with spectroscopic data
allows for improved photo-z and shear calibration and IA mitigation
schemes.

Another interesting prospect is the design of an optimal TF-lensing
tomographic survey. The small shape noise and the accurate redshift
information allows for substantially more tomographic bins and hence
for a precise measurement of expansion history vs structure growth. We
point out that the TF-lensing method presented in this paper can be
applied to cluster lensing, galaxy-galaxy lensing and other
cross-lensing probes, thereby overcoming possible limitations of these
probes due to shape noise.

Using the TF method presented in this paper cosmic shear for the first
time is no longer fundamentally limited by shape noise errors and
systematics associated with it but by the instrumental capabilities of
multi-object spectrographs.

\section*{Acknowledgments}
We thank Klaus Honscheid, David Weinberg, Rachel Mandelbaum, Chris Hirata, 
Bhuvnesh Jain, and Peter Schneider for very useful discussions and advice. 
We also thank the Center for Cosmology and Astroparticle Physics at the Ohio State University and 
the Center for Particle Cosmology at the University of Pennsylvania for hosting
us during critical phases of the paper. This paper is based upon work
supported in part by the National Science Foundation under Grant
No. 1066293 and the hospitality of the Aspen Center for Physics. This
research funded in part by by NASA ROSES ATP 17-ATP17-0173.

\bibliographystyle{aasjournal}
\bibliography{refs}

\end{document}